\documentstyle[12pt]{article}
\begin{document}

\begin{flushright}
                   {\small gr--qc/9407022}\\
\end{flushright}
\vskip .5cm
\centerline{\bf \LARGE A Note on Hartle-Hawking Vacua}
\vskip 1cm
\centerline{Ted Jacobson\footnote{jacobson@umdhep.umd.edu}}
\vskip .2cm
\centerline{\it Dept. of Physics,
Univ. of Maryland, Collge Park, MD 20742-4111}
\vskip 1cm

\begin{abstract}
{The purpose of this note is to  establish the basic
properties---regularity at the horizon, time independence,
and thermality---of the
generalized Hartle-Hawking vacua defined in static spacetimes
with bifurcate Killing horizon admitting a regular Euclidean section.
These states, for free or interacting fields, are defined by a
Euclidean functional integral on half the Euclidean section.
The emphasis is on generality  and the arguments are simple
but formal.}
\end{abstract}

The Hartle-Hawking vacuum\cite{HH} was originally introduced
via the assumption that the Feynman propagator for a free field
in Schwarzschild spacetime is given by
analytic continuation from the 2-point function on the Euclidean
section. Due to the periodicity in
time of the Euclidean section of the black hole spacetime,
this assumption yields
a 2-point function on the Lorentzian section that has thermal
properties. It was soon realized\cite{Israel}
that, at least in the case of free fields,
the underlying state is a pure state that involves correlations
between modes localized on either side of the horizon.
It was later argued\cite{GP} more generally using perturbation theory
that, on account of periodicity of the 2-point function,
all n-point
correlation functions for an arbitrary interacting quantum field
on the Euclidean section are thermal correlation functions
when continued back to the Lorentzian section.

The well known
equivalence between the Minkowski vacuum, and a thermal
state defined by the Rindler hamiltonian generating evolution
along a boost Killing field in Minkowski spacetime, can easily be
seen using a path integral representation for the wave
functional of the vacuum.
[Several authors have, apparently independently, pointed this out
\cite{UW,LaFl,CW,KS}.]
This path integral representation, which applies
to interacting fields as well as free fields,
can be generalized to arbitrary static spacetimes
with bifurcate Killing horizons such as
static black holes and deSitter space.
Indeed, this generalization was pointed out in \cite{LaFl},
and it was recently shown\cite{Barv} by explicit computation in
the free field case that the path integral in
Euclidean Schwarzschild spacetime yields the Hartle-Hawking vacuum.

The purpose of this note is to establish the basic  properties
of the generalized Hartle-Hawking ``vacuum" states.
The emphasis here is on generality.
That the states are invariant under the Killing field and regular at
the event horizon will be shown by simple formal arguments, as will be
their precisely thermal nature as viewed
from one side of the horizon.
[``Thermality" is defined here with respect to the
Hamiltonian that generates evolution along the static Killing field,
and the temperature is $\kappa/2\pi$, where $\kappa$ is the
surface gravity of the Killing horizon.]

Consider a quantum field $\phi$ propagating in a static spacetime
with a bifurcate Killing horizon with a regular bifurcation surface
and (therefore\cite{RaczWald}) constant surface gravity.
It is known that for a free field,
there is at most one Killing time independent quantum state
$|0\rangle$ that is regular everywhere on the horizon\cite{KW}.
For example, in Schwarzschild spacetime, this state is
the Hartle-Hawking ``vacuum". In the interacting
case there might be multiple vacua, but the path integral will
select just one of them, which will be referred to as $|0\rangle$.

Consider a spacelike hypersurface $\Sigma$, orthogonal to the
Killing field, which includes the bifurcation surface $B$.
The bifurcation surface divides $\Sigma$ into two pieces,
$\Sigma_I$ and $\Sigma_{II}$, whose intersection is $B$. The
``Euclidean section" is the Riemannian 4-D ``surface of revolution"
obtained by
rotating each point of $\Sigma_I$ through a circle of radius
$N\beta$. Here $N$ is the norm of the Killing field at the point,
and $\beta$ is the
period of the Euclidean time coordinate required in order that
there be no conical singularity at the (Euclidean) horizon
(where $N=0$).
Alternatively, $\beta=2\pi/\kappa$, where
$\kappa=\lim_{N\to 0}|\nabla_a N|$ is the surface gravity of the
Killing horizon.
The points of $B$ are fixed points of the Killing field.

The nature of the surface $\Sigma_I$ depends on the spacetime.
For instance, in Minkowski space with a boost Killing field
$B$ is $R^2$ and $\Sigma_I$ is $R^2\times R^+$. In
Schwarzschild space $B$ is a two-sphere and $\Sigma_I$ is
topologically $S^2\times R^+$. In deSitter space $B$ is also
an $S^2$, the equator of $S^3$, and $\Sigma_I$ is the hemisphere
$D^3$. Note that in the first two cases $\Sigma_I$ and the
Euclidean section are non-compact,
whereas in deSitter space they are compact.

The state $|0\rangle$ is defined in the functional Schr\"odinger
representation as a path integral over fields living on half
of the Euclidean section of the spacetime. The action is
the Euclidean one, and the fields are assumed to
take a fixed value on the boundary surface
consisting of the union of the two Euclidean Killing times
$\tau=0$ and $\tau=\beta/2$.
If the Euclidean section is noncompact, the fields are assumed to
vanish asymptotically to ensure that the path integral is
well defined.

The functional defined by this path
integral also defines a state on the surface
$\Sigma=\Sigma_I\cup\Sigma_{II}$ in the Lorentzian spacetime
when one identifies, say, $\Sigma_{II}$
as the $\tau=0$ Euclidean time slice and $\Sigma_I$ as the
$\tau=\beta/2$
time slice.
This state is regular everywhere on $\Sigma$ provided
that we choose the range of $\tau$ on the
half-space to be precisely $\Delta\tau=\beta/2$.
In fact, the $n$-point correlation functions in the state
$|0\rangle$ with all points on $\Sigma$ are just those defined in the
field theory on the Euclidean section of period $2\Delta\tau$.
If $\Delta\tau\ne\beta/2$, this Euclidean section will  have
a conical singularity, and the correlation functions will not
have the usual short distance behavior when one or more of
the points sits at the conical singularity.
Thus the state will not be regular at the horizon.
If, on the other hand, $\Delta\tau=\beta/2$, then the Euclidean section
is regular everywhere and therefore so is the state.

The state $|0\rangle$ is invariant under the time translation
Killing field that vanishes at the bifurcation surface.
To establish time independence, let $H$ be the Hamiltonian
generating translations in Euclidean Killing time $\tau$.
(Subtleties associated with the boundary of the spatial slices
at $B$ are ignored in what follows.)
Then the path
integral we have defined just gives the matrix elements of the
operator ${\cal N}\exp(-\frac{\beta}{2}H)$,
where ${\cal N}$ is a normalization factor.
Identifying the Hilbert space ${\cal H} _{II}$
associated with the field on $\Sigma_{II}$ with the dual of
that (${\cal H}_I$) associated with the field on $\Sigma_{I}$, the operator
$\exp(-\frac{\beta}{2}H)$ becomes a vector in the tensor product
${\cal H}_I\otimes{\cal H}_{II}$.
Since $\Sigma$ is orthogonal to the Killing field,
 the Lorentzian and Euclidean hamiltonians
generating evolution along the Killing flow agree on $\Sigma$.
More precisely, the Lorentzian hamiltonian is equal to $H$ on
$\Sigma_I$, and equal to $-H$ on $\Sigma_{II}$,
since Killing time flows ``backward" in region $II$.
The action of the
Lorentzian hamiltonian on $|0\rangle$ is therefore
given by the commutator
$[H,\exp(-\frac{\beta}{2}H)]=0$, so the state $|0\rangle$ is indeed
time-independent.

The sense in which $|0\rangle$ represents a thermal state is
manifest when it is expressed as the operator
$|0\rangle={\cal N}\exp(-\frac{\beta}{2}H)$.
Tracing over the field degrees of freedom in region $II$
we obtain the reduced density matrix
$\rho_I\equiv {\rm Tr}_{II}|0\rangle\langle0|={\cal N}^2\exp(-\beta H)$,
which describes a {\it mixed}
state rather than a pure one, due to the entanglement of
the degrees of freedom in regions $I$ and $II$ in the state
$|0\rangle$.
$\rho_I$ is precisely the density matrix
for the canonical thermal ensemble
at the temperature $\beta^{-1}$, defined with respect to the
hamiltonian of the static observers in region $I$ outside the
Killing horizon. Note that the ``entanglement entropy"
$-{\rm Tr}\rho_I\ln\rho_I$ is thus identical to the thermal entropy
of the canonical ensemble.

This completes the task of establishing the basic
properties---regularity at the horizon, time independence,
and thermality---of the
generalized Hartle-Hawking vacua defined in static spacetimes
with bifurcate Killing horizon admitting a regular Euclidean section.
These states are defined by a Euclidean functional integral on half
the Euclidean section or, equivalently, by the operator
$\exp(-\frac{\beta}{2}H)$. We conclude this note with three remarks.

($i$) The state $|0\rangle$ may become singular after Lorentzian time
evolution off the surface $\Sigma$. For example in Reissner-Nordstr\"om
spacetime, as is well known, the state becomes singular at the inner
(Cauchy) horizon.

($ii$) The construction employed here does not exist for the extremal
Reissner-Nordstr\"om black hole, since that spacetime has no
bifurcation surface. It also does not exist for the
Schwarzschild-deSitter spacetime, except when the cosmological and
event horizons have the same surface gravity, since a difference
in surface gravities precludes the existence of a regular Euclidean
section.

($iii$) In the form given here the construction does not apply to
stationary nonstatic spacetimes because there is no Euclidean section.
However, it seems likely that an appropriate generalization could be
found.
\vskip 1cm
\begin{flushleft}
{\Large \bf  Acknowledgments}
\end{flushleft}

I am grateful to Jonathan Simon for useful discussions and criticism.
I also thank the Institute for Theoretical Physics at the University
of Bern for hospitality and support while this work was being completed.
This work  was supported in part by NSF grant PHY91-12240.

\end{document}